\documentclass[aps,prb,twocolumn,showpacs,superscriptaddress]{revtex4}

\usepackage{amsmath}
\usepackage{epsfig}
\usepackage{bm}
\usepackage{longtable}
\usepackage{graphicx}
\usepackage{graphics}

\begin{document}

\title{Josephson oscillations between exciton
condensates in electrostatic traps
}

\author{Massimo Rontani}
\affiliation{CNR-INFM Research Center on nanoStructures
and bioSystems at Surfaces (S3), 
Via Campi 213/A, 41125 Modena, Italy}
\email{rontani@unimore.it}
\homepage{www.nanoscience.unimore.it/max.html}

\author{L. J. Sham}

\affiliation{Department of Physics, University of California San Diego,
La Jolla, California 92093-0319}

\date{\today}

\begin{abstract}
Technological advances allow for tunable lateral confinement of
cold dipolar excitons in coupled quantum wells.
We consider theoretically the Josephson effect between exciton
condensates in two traps separated by a weak link.
The flow of the exciton supercurrent is driven by the 
dipole energy difference between the traps.
The Josephson oscillations may be observed
after ensemble average of the time correlation
of photons separately emitted from the two traps.
The fringe visibility  
is controlled by the trap coupling and is
robust against quantum and thermal fluctuations.
\end{abstract}


\maketitle

\section{Introduction}

The Josephson effect is a a macroscopic coherent phenomenon which has been 
observed in systems as diverse as superconductors,\cite{Barone} 
superfluid Helium,\cite{Helium} Bose-Einstein condensates in 
trapped ultra-cold atomic gases.\cite{Atomic}
Since Josephson oscillations appear naturally when 
two spatially separated macroscopic wave functions are weakly coupled,
they have been predicted for  bosonic excitations in solids as well, like 
polaritons \cite{Carusotto,Sarchi} and excitons.\cite{Shelykh}
However, unlike the polaritons, which have a photonic component 
allowing for easy detection,\cite{Littlewood} excitons stay dark unless they
recombine radiatively. So far, it is unclear how the exciton
Josephson effect could be observed. In this paper, we
propose an experiment. 

Condensed excitons are predicted to emit coherent 
light.\cite{Oestreich96} If Josephson oscillations occur
between two exciton traps, in principle they can be probed by 
measuring the interference of the beams separately emitted from the traps. 
However, in the time interval before recombination, 
there are too few photons emitted for an adequate signal to noise ratio,
and one has to average the signal over many replicas of the same 
experiment.\cite{Yang06}
We will show that such ensemble averaging
blurs the signature of the Josephson effect 
except in the relevant case of exciton ``plasma'' 
oscillations.\cite{Pitaevskii} For the latter the
dipole energy difference between the traps modulates 
the visibility $\alpha$
of interference fringes, providing a means for detection. 

The paper is organized as follows:
In Sec.~\ref{s:electrical} we introduce the double quantum well system
and illustrate a feasible scheme to manipulate electrically
the exciton phase. After 
setting the theoretical framework 
(Sec.~\ref{s:Josephson}), we discuss the proposed correlated photon
counting experiment (Sec.~\ref{s:correlated}) and provide
an estimate for its key parameters (Sec.~\ref{s:estimate}).

\section{Electrical control of the exciton phase}\label{s:electrical}

\begin{figure}
\setlength{\unitlength}{1 cm}
\begin{picture}(8.5,2)
\put(0.1,0.2){\epsfig{file=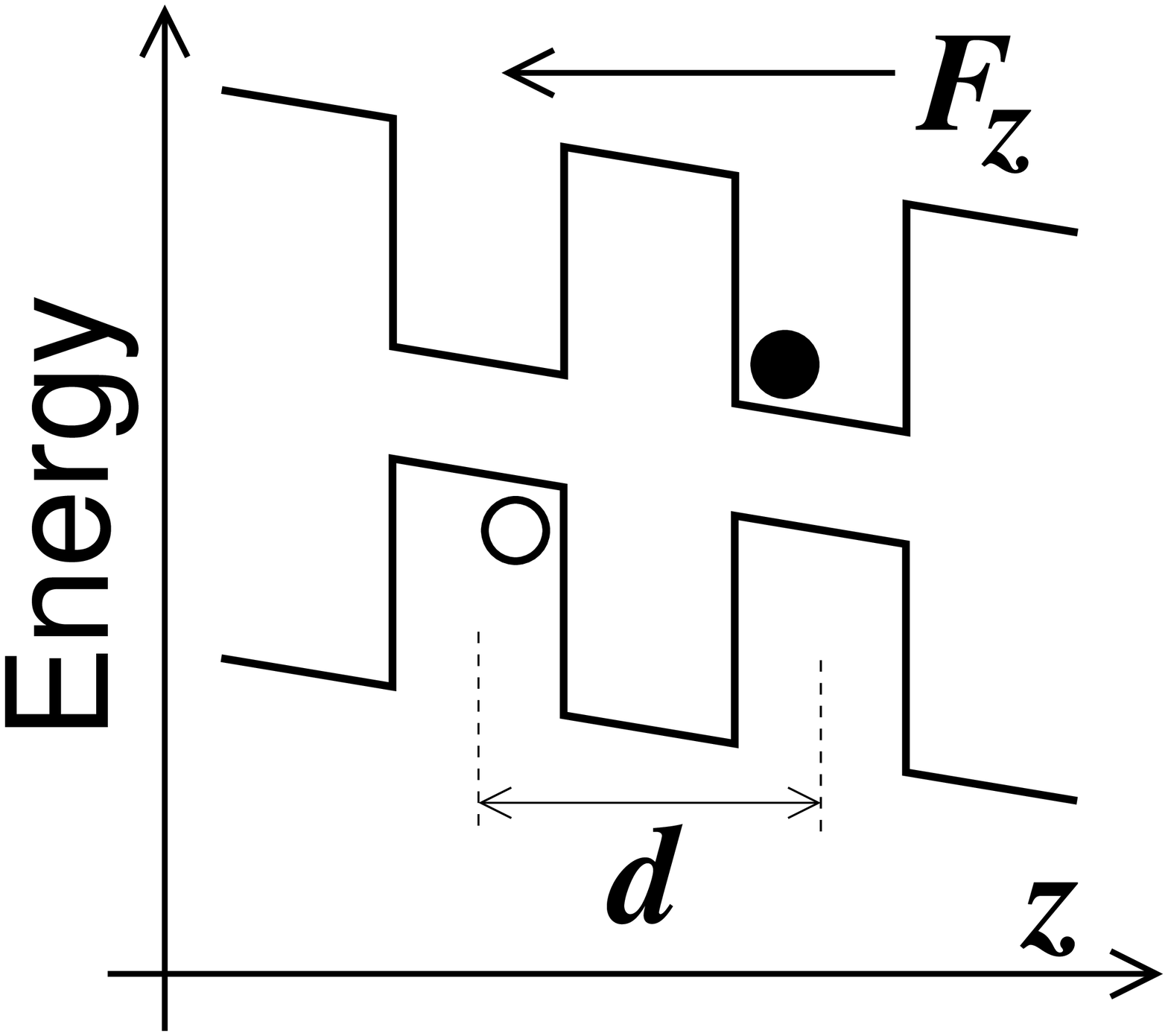,angle=-0,width=1.7cm}}
\put(2.2,0.2){\epsfig{file=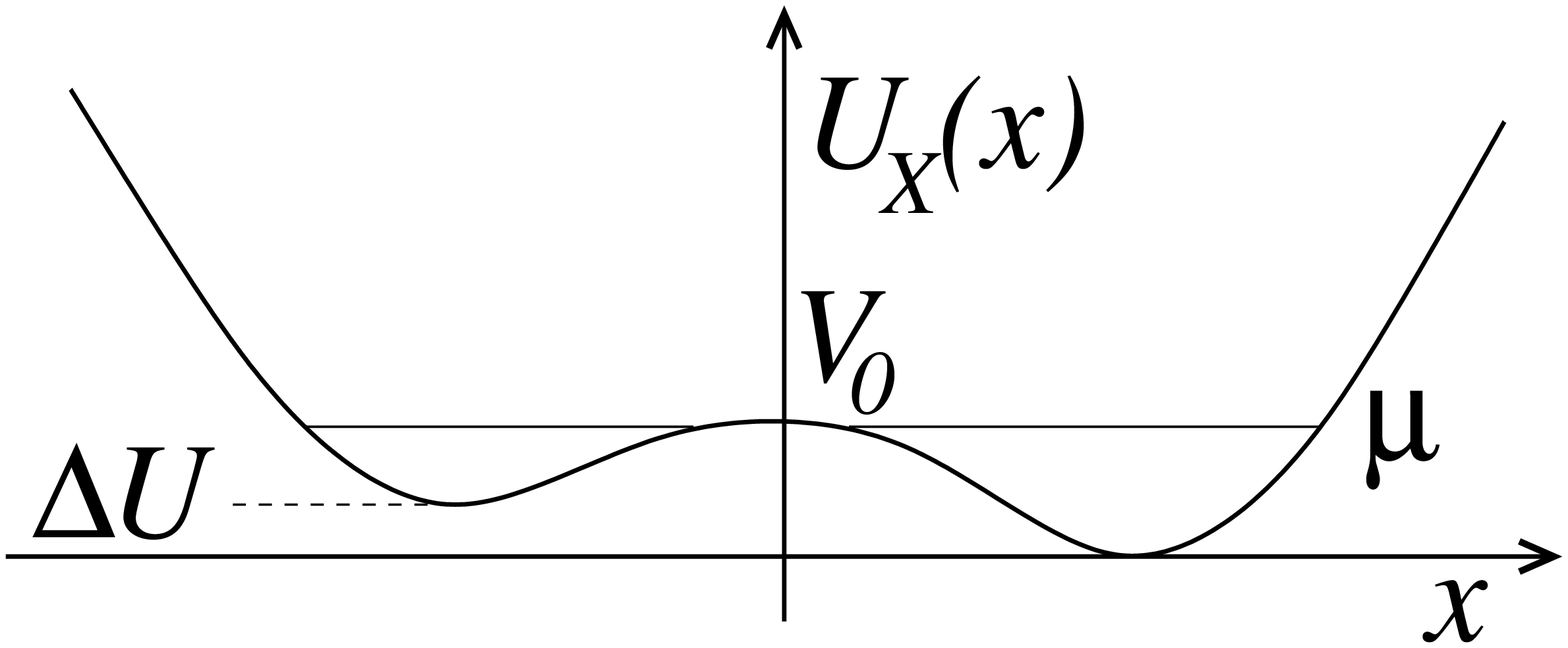,angle=-0,width=3.2cm}}
\put(5.7,0.0){\epsfig{file=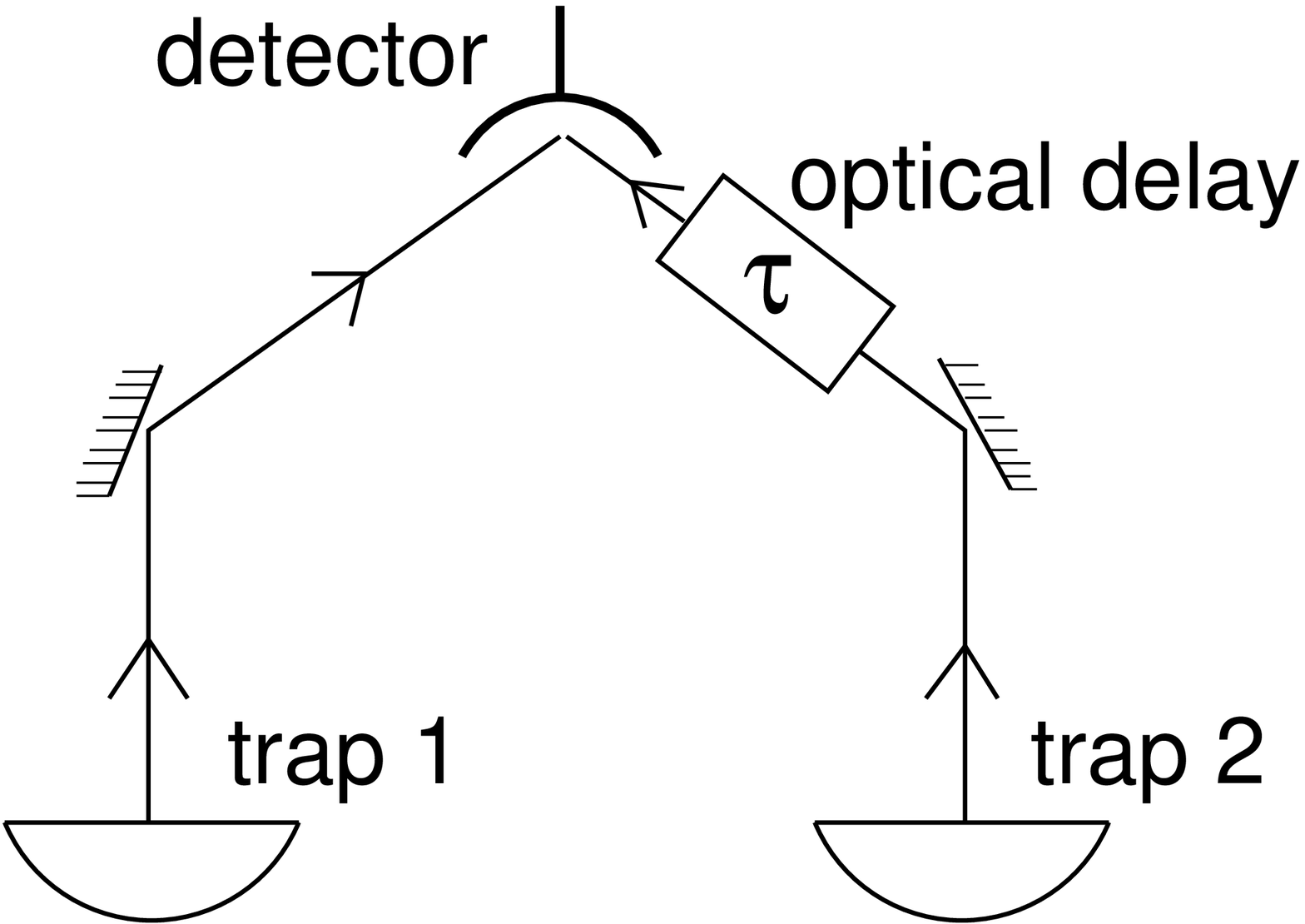,angle=-0,width=2.5cm}}
\put(1.8,1.6){(a)}
\put(8.0,1.6){(c)}
\put(5.0,1.6){(b)}
\end{picture}
\caption{(a) Double quantum well energy profile along the growth
direction $z$.
(b) Exciton potential profile of the double-trap system along $x$.
(c) Arrangement for measuring the time correlation of the emitted 
photons from the two traps.
}
\label{fig1}
\end{figure}

Consider a double quantum well where electrons and holes 
are separately confined in the two layers. In experiments
aiming at Bose-Einstein condensation of excitons, 
electron-hole pairs are optically generated off-resonance, 
left to thermalize, form excitons, 
and, at sufficiently low temperature $T$ and high density,
condense before radiative decay.\cite{Butov-long}
Let $z$ be the growth axis of the two wells separated by 
distance $d$. The electrons in the conduction band and 
holes in the valence band move in the planes $z=d$, $0$, 
respectively [Fig.~\ref{fig1}(a)]. 
Let $\Psi_a(x,y,0,t)$ and $\Psi^{\dagger}_b(x,y,d,t)$ denote the respective 
electron and hole destruction operators, with the vacuum being the 
semiconductor ground state with no excitons. 
In the experiments,\cite{Butov-long} 
an electric field $F_z$ is applied along $z$ 
to suppress inter-layer tunneling, thereby 
quenching the exciton recombination.
Fabrications \cite{Hammack} of electrostatic traps with suitably located 
electrodes to provide lateral confinement for the 
excitons have been implemented. 
The double quantum well is sandwiched between two spacer layers, 
providing insulation from planar electrodes lithographed on both sides of 
the coupled structure.
Each electrode controls a tunable gate voltage, $V_g(x,y,z)$, which 
localizes in a region of the $xy$ plane the field component 
along $z$, $F_z(x,y,z)=-\partial V_g(x,y,z) /\partial z$, 
while $F_x$ and $F_y$ are 
small and can be neglected as well as the the dependence of $F_z$ on $z$. 
The vertical field $F_z(x,y)$  makes the electrostatic potential energy 
of the exciton dipole 
depend on the lateral position, $U_X(x,y)=-edF_z(x,y)$ ($e<0$)
[cf.~Fig.~\ref{fig1}(b)].
In this way, potential traps for excitons are 
designed with great flexibily, with
 {\em in situ} control of the height, width, and shape of the 
potential barriers.\cite{Hammack}

First, we focus on the quasi-equilibrium situation before radiative 
recombination, where excitons  condense in two coupled electrostatic 
traps, both within the condensate coherence length.
Figure \ref{fig1}(b) depicts the exciton potential 
profile $U_X(x,y=0)$ along the $x$ axis, with a link between two identical 
traps. The potential barrier allows tunneling between the 
condensates $\Xi_1(x,y,t)$ and $\Xi_2(x,y,t)$ in the two 
traps.\cite{RontaniPRL} The optical coherence in a single trap is of the form
\begin{equation}
\Xi(x,y,t) = \left< \Psi^{\dagger}_a(x,y,0,t) \Psi_b(x,y,d,t) \right>\;,
\label{eq:definition}
\end{equation}
where $\left<\ldots\right>$ denotes quantum and thermal average. 
In the limit $n\, a_{\text{B}}^2
\ll 1$, with $a_{\text{B}}$ being the two-dimensional
effective Bohr radius and $n$ the exciton density, 
$\Xi(x,y,t)$ is the macroscopic wave function
for the center-of-mass motion of excitons,\cite{Keldysh}
which may be written in the form
\begin{equation}
\Xi(x,y,t) = \sqrt{n_s}\; e^{{\rm i}\varphi},
\end{equation}
with $n_s$ being the density of the exciton condensate and $\varphi$ the
phase.\cite{SBGS} 
For a gauge transformation of the gate potential 
$V_g\rightarrow V_g -c^{-1}\partial \chi(t)/\partial t $,
which leaves the field $F_z$ unaltered, 
the field operators $\Psi$ gain a phase,
\begin{eqnarray}
\Psi_{a} &\rightarrow & 
\Psi_{a} \,\exp{\!\left[\!\frac{{\rm i}e}{\hbar c}\,\chi(x,y,0,t)\right]}\;, 
\nonumber\\ 
\Psi_{b} & \rightarrow & \Psi_{b} \,\exp{\!\left[\frac{{\rm i}e}{\hbar c}
\,\chi(x,y,d,t)\right]}\; .
\end{eqnarray}
The macroscopic wave function, by Eq.~(\ref{eq:definition}), 
also gains a phase,
\begin{equation}
\varphi \rightarrow \varphi + \frac{e}{\hbar c}\left[\chi(z=d,t)
-\chi(z=0,t)\right]\;.
\label{eq:phitransf}
\end{equation}
Hence, the frequency of time oscillation of the condensate 
is given by the electrostatic energy of the exciton dipole
in the external field,\cite{Balatsky} $U=-edF_z$:
\begin{equation}
\varphi = \varphi^{(0)} + \frac{1}{\hbar}edF_zt\;,
\label{eq:phitime}
\end{equation} 
with $\varphi^{(0)}$ being 
the time-independent zero-field value.\cite{phase}
In the absence of the bilayer separation of the electrons 
and the holes, their gauge phases 
gained in the electric field would cancel each other
resulting in no time dependence driven by $U$.
Equation (\ref{eq:phitime}) shows that the experimentally controllable 
dipole energy difference between the two traps depicted 
in Fig.~\ref{fig1}(b),  $\Delta U= -ed(F_{z1}-F_{z2})$, 
drives the relative phase between the two condensates, 
thereby creating Josephson oscillations as a means for 
measuring the Josephson tunnel between the traps. 

\section{Exciton Josephson oscillations}\label{s:Josephson}

We next introduce the usual two-mode description of inter-trap
dynamics based on the Gross-Pitaevskii (GP)
equation.\cite{Carusotto,Sarchi,Shelykh,Smerzi,Zapata,Pitaevskii}
Exciton-exciton correlation 
\cite{Oestreich} beyond the GP mean field  
may be neglected due to the repulsive character of
the dipolar interaction 
between excitons in coupled quantum wells.
The condensate total wave function solution is
\begin{equation}
\Xi(x,y,t) = \Xi_1(x,y,N_1)\,\text{e}^{i\varphi_1} +
\Xi_2(x,y,N_2)\,\text{e}^{i\varphi_2}\;,
\label{eq:two}
\end{equation}
where both the trap population $N_i(t)$ and the condensate phase 
$\varphi_i(t)$ 
possess the entire time dependence for the $i$th trap  ($i=1,2$),  and 
$\Xi_i(x,y,N_i)$ is a real quantity, with 
\begin{equation}
\int\!\! dx\!\!\int\!\! dy\; \Xi_i^2(x,y,N_i) = N_i(t).
\end{equation} 
The dynamics
of the GP macroscopic wave function
$\Xi(x,y,t)$ depends entirely on
the temporal evolution of two variables, the population imbalance
$k(t)=(N_1-N_2)/2$ and the relative phase $\phi(t)=\varphi_1-\varphi_2$
of the two condensates. Here we  consider a
time interval much shorter than the exciton lifetime
(10 --- 100 ns) and ignore the spin structure.
Therefore, the total population is approximately constant,
$N_1(t)+N_2(t)=N$. The equations of motion for the
canonically coniugated variables $\hbar k$ and $\phi$
are derived from the effective Hamiltonian
\begin{equation}
H_J = E_c\frac{k^2}{2} +\Delta U k
-\frac{\delta_J}{2}\sqrt{N^2-4k^2}\cos{\phi}\;,
\label{H}
\end{equation}
under the condition $k\ll N$ (Ref.~\onlinecite{Pitaevskii}).
$E_c=2d\mu_1/dN_1$ is the exciton ``charging'' energy of one trap, 
where $\mu_1$ is the chemical potential of trap 1,
whereas $\delta_J$ is the Bardeen single-particle tunnelling energy,
\begin{equation}
\delta_J=\frac{\hbar^2}{m}\int\!\!dy\left[\xi_1\!
\left(\frac{\partial \xi_2}{\partial x}\right)
-\xi_2\!\left(\frac{\partial \xi_1}{\partial x}\right)\right]_{x=0},
\end{equation}
where $m$ is the exciton mass. The single-particle orbital $\xi_i(x,y)$
is defined through $\Xi_i(x,y)=\sqrt{N_i}\,\xi_i(x,y)$.

The various dynamical regimes associated to certain
intitial conditions $\left(k(0),\phi(0)\right)$,
including $\pi$ oscillations
and macroscopic quantum self-trapping, are 
exhaustively discussed in Refs.~\onlinecite{Smerzi}. 
Two cases are specially relevant:

\subsubsection{AC Josephson effect} 

Under the conditions $\Delta U \gg NE_c/2$,
$\Delta U\gg \delta_J$, one easily obtains
\begin{equation}
\phi(t) = -\frac{\Delta U}{\hbar}t+\phi(0)\;,\qquad 
\dot{k} =\frac{\delta_JN}{2\hbar}
\sin{\phi}\;.
\label{AC}
\end{equation}
Equation (\ref{AC}) shows that, analogously to the 
case of two superconductors separated by a thin barrier, if the
phase difference $\phi$ between the condensates is not a multiple
of $\pi$, an exciton supercurrent $2\dot{k}$ flows across the barrier. 
Remarkably, in the presence of an electric field gradient along $z$,
an exciton flux oscillates back and forth between the two traps,
with frequency $\Delta U/\hbar$. 
As an exciton goes through the barrier, it exchanges with
the field the dipole energy acquired or lost in the 
tunneling process.
The analogy with the AC Josephson effect for
superconductors is clear: in that case 
a bias voltage $V$ is applied across the junction,
and the energy $2eV$ is exchanged between field and 
Cooper pairs, as the latter experience a potential
difference of $V$ when penetrating the potential barrier.

\subsubsection{Plasma oscillations}

This case concerns small oscillations around
the equilibrium position $(k,\phi)_{\text{eq}}=(0,0)$. 
The Hamiltonian (\ref{H}) may then be linearized into the form
\begin{equation}
H_J = \frac{k^2}{2}\left(2\frac{\delta_J}{N}+E_c\right)+\frac{1}{4}\delta_J
N\phi^2+\Delta U k-\frac{\delta_JN}{2}\;.
\label{Hnormal}
\end{equation}
It follows that both $k$ and $\phi$ oscillate in time 
with plasma frequency 
\begin{equation}
\omega_J=\frac{1}{\hbar}\sqrt{\delta_J(NE_c/2+\delta_J)}. 
\end{equation}
Note that $\Delta U$ displaces the equilibrium position
from $(k,\phi)_{\text{eq}}=(0,0)$ to 
\begin{equation}
(k,\phi)_{\text{eq}}=(-\Delta U N \delta_J/2(\hbar\omega_J)^2,0).  
\label{eq:displace}
\end{equation}

\section{Correlated photon counting experiment}\label{s:correlated}

Figure \ref{fig1}(c) illustrates 
the correlated photon counting setup which
we propose to probe Josephson oscillations. The detector
measures the intensity $I(\tau)$ 
of the sum of the two beams separately emitted from
the traps. A delay time $\tau$ is induced in one of the two
beams, as in Ref.~\onlinecite{Yang06}. 
The fields are simply proportional
to the order parameters $\Xi_i$ of the traps. In fact, $\Xi(x,y,t)$
is associated with a macroscopic
electric dipole
moment, $\bm{P}(t)=\hat{\bm{x}}P_x(t)\pm i\hat{\bm{y}}P_y(t)$, which 
couples to photons:
$P_{x}(t)=\int dx\,dy \, x\,\, \Xi(x,y,t)$,
and similarly for $P_y$.
The built-in dipole 
$\left<\bm{P}(t)\right>\neq 0$ oscillates
with frequency $(\mu+E_X)/\hbar$,
where $E_X$ is the optical gap minus the exciton binding energy,
and $\mu$ accounts for exciton-exciton interaction.\cite{Oestreich96} 
This macroscopic oscillating 
dipole is equivalent to a noiseless current, which 
radiates a coherent field.\cite{Glauber}
Therefore, the measured intensity $I(\tau)$ is
$I(\tau)=2I_0\left[1+\left<\cos{\phi(\tau)}\right> \right]$,
assuming that the fields emitted from the two traps have 
the same magnitude (and intensity $I_0$) 
but different relative phase $\phi$, which is evaluated 
at the delayed time $\tau$.\cite{notespectral} $I(\tau)$ may be written as
\begin{equation}
I(\tau)=2I_0\left[1+\alpha\cos{\phi_0(\tau)} \right]\;,
\label{interfe2}
\end{equation}
where $\phi_0(\tau)$ is the phase averaged over many measurements, defined
by the condition $\left<\sin{[\phi(\tau)-\phi_0(\tau)]}\right>=0$,
and 
\begin{equation}
\alpha=\left<\cos{[\phi(\tau)-\phi_0(\tau)]}\right>
\end{equation}
is the fringe visibility,   
i.e., the normalized peak-to-valley ratio of fringes,
$\alpha=(I_{\text{max}}-I_{\text{min}})
/(I_{\text{max}}+I_{\text{min}})$, 
with $I_{\text{max}}$ ($I_{\text{min}}$) being
the maximum (minimum) value of $I(\tau)$, and $0\le\alpha\le 1$.

Equation (\ref{interfe2}) has a few important caveats.
Since $I(\tau)$ is an average,
the temporal inhomogeneous effect will 
blur the interference fringes, i.e., $\alpha<1$.
Other dephasing mechanisms include exciton recombination
and inelastic exciton-phonon scattering,\cite{Yang06}  as well as 
inelastic\cite{Yang06} and elastic\cite{Zimmermann}
exciton-exciton scattering, 
which in first instance may all be neglected for
short $\tau$, low $T$, and $n\, a_{\text{B}}^2 \ll 1$, respectively. 
The most immediate caveat is that the exciton condensates 
in decoupled traps must acquire a relative phase if initially 
they condense separately without a definite phase relation.
This scenario is analogous to the case of interference
between independent laser sources first discussed by Glauber 
\cite{Glauber} and later studied experimentally 
for matter waves.\cite{Andrews}
Even though a one-shot measurement with sufficient resolution 
would display fringes, the relative phase $\phi_0(\tau)$ is also 
subject to intrinsic dephasing effects by
quantum fluctuations.\cite{Glauber}
The latter are significant noise sources which affect $\alpha$, 
when $\phi$ and $k$ are quantized into canonically
conjugated quantum variables whereas 
in the GP theory used so far they were 
classical variables whose fluctuations where neglected.

In the following, we quantize Hamiltonian (\ref{H})
in order to properly evaluate 
$\alpha=\left<\cos{[\phi(\tau)-\phi_0(\tau)]}\right>$ 
as a quantum statistical average in finite traps.
Therefore, we follow Ref.~\onlinecite{Pita} 
and introduce the commutator
$[\hat{\phi},\hat{k}]=i$. 
The operator $\hat{k}$ now appearing in the quantized
version of Hamiltonian (\ref{H}) takes
the form $-i\partial / \partial \phi$,
whereas
the ground state wave function
is defined in the space of periodical functions of $\phi$ 
with period $2\pi$.
If condensate oscillations are mainly coherent, 
the variance of $\phi$ is small and the visibility is 
approximated by $\alpha= 1- \frac{1}{2} \left<(\Delta\phi)^2\right>$. 

The most interesting case concerns plasma oscillations. 
For $\Delta U = 0$, the ground state of the quantized version
of the harmonic
oscillator Hamiltonian (\ref{Hnormal}) is a Gaussian, 
with $\phi_0=0$, independent from $\tau$,
and minimal spreading $\left<\Delta\phi^2\right>\approx 
(E_c/2\delta_J N)^{1/2}$. Therefore, the 
interferometer output is time-independent,
$I=2I_0(1+\alpha)$, showing constructive interference,
$I>2I_0$,
with $\alpha=1-(E_c/8\delta_J N)^{1/2}$.
Not surprisingly, the visibility is controlled by
the ratio $E_c/\delta_J N$, reaching the maximum $\alpha =1$
as $ E_c/\delta_J N \rightarrow 0$. In fact, $\alpha$ is
given by the balance between the competing effects of
tunnelling ($\propto \delta_JN$), which enforces a 
well-defined inter-trap phase, and inverse 
compressibility ($\propto E_c$), which favors the formation
of separate number states in the two traps, 
thus separating the two macroscopic wave functions.

\begin{figure}
\includegraphics[width=3.0in]{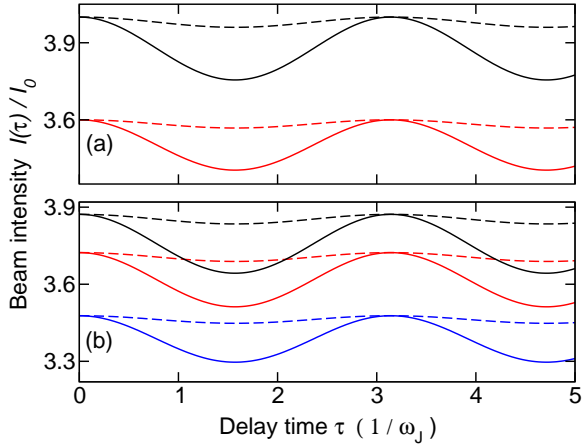}
\caption{(Color online).  Beam intensity $I(\tau)/I_0$
vs delay time $\tau$, for $\Delta U/\hbar\omega_J=$ 0.2, 0.5 (dashed 
and solid lines, respectively). (a) $T=$ 0 and $\alpha=$ 1, 0.8
(black and red [light gray] lines, respectively). (b) $\alpha(T=0)=$ 0.94 and 
$k_B T/\hbar\omega_J = $ 0, 1, 2 (black, red [light gray], and blue 
[dark gray] lines, respectively).
\label{fig2}}
\end{figure}

A small finite value of $\Delta U$ in Eq.~(\ref{Hnormal})
displaces the equilibrium position
of the harmonic oscillator. Noticeably, the ground state is 
a \emph{coherent} state with non-null evolution of the average phase
in time, 
\begin{equation}
\phi_0(\tau)=-\frac{\Delta U}{\hbar\omega_J}\sin{(\omega_J\, \tau)},
\end{equation}
whereas $\alpha$ is unchanged. This key feature allows for
directly monitoring $\tau$-dependent 
plasma oscillations of frequency $\omega_J$
through the photon correlation measurement
(cf.~Fig.~\ref{fig2}).
We evaluate the effect of thermal fluctuations on 
$\alpha$ via the formula
$\alpha(T) = \sum_n \alpha_n \exp{[-\beta E_n]}/ \sum_n \exp{[-\beta E_n]}$,
where $\beta = 1/k_B T$, $k_B$ is the Boltzmann
constant, $2(\alpha_n-1)=\left<(\Delta\phi)^2\right>_n$ is the
variance of $\phi$ in 
the $n$th excited state whose energy is $E_n$. At low $T$, 
the excited states may be approximated as those
of the harmonic oscillator, giving 
\begin{equation}
\alpha(T)=1-\sqrt{\frac{E_c}{2\delta_J N}}
\left(\frac{1}{2}+\frac{1}{e^{\beta \hbar \omega_J}
-1}\right).
\end{equation}

The above results are summarized by the formula
\begin{equation}
I(\tau)=2I_0\!\left[1+\alpha(T)\cos{\!\left(\frac{\Delta U}{\hbar
\omega_J}\sin{\omega_J\,\tau}\right)} \right],
\label{interfe3}
\end{equation}
which is valid for $E_c/\delta_J N \ll 1$. 
For small dipole energy variations, $\Delta U/ \hbar \omega_J \ll 1$,
the oscillating part within the square brackets of Eq.~(\ref{interfe3})
may be written as 
$-\alpha(T)/2(\Delta U/\hbar \omega_J)^2\sin^2{\omega_J\,\tau}$.
This shows that the visibility $\alpha(T)$ of fringes, which oscillate
like $\sin^2{\omega_J\,\tau}$, is modulated by
the experimentally tunable factor $(\Delta U/\hbar \omega_J)^2/2$.
The dependence of $I(\tau)$ on $\Delta U$ 
is illustrated in Fig.~\ref{fig2} for two values of
$\Delta U/\hbar\omega_J$. As $\Delta U/\hbar\omega_J$ is increased
[from 0.2 (dashed lines) to 0.5 (solid lines)], the amplitude of
oscillations of $I(\tau)$ shows a strong non-linear enhancement, 
providing a clear signature of Josephson oscillations.  
The oscillation amplitudes are 
larger for higher values of $\alpha$ [cf.~Fig.~\ref{fig2}(a)], 
and fairly robust against thermal smearing 
[cf.~Fig.~\ref{fig2}(b)]. In fact, Fig.~\ref{fig2}(b) shows
that the oscillation of $I(\tau)$ is still clearly resolved 
for temperatures as high as
$T\approx \hbar\omega_J/k_B$. At even higher temperatures
$\alpha(T)$ displays anharmonic effects,\cite{Pita} with
$\alpha(T)\rightarrow 0$ as $T\rightarrow \infty$.

\begin{figure}
\includegraphics[width=3.0in]{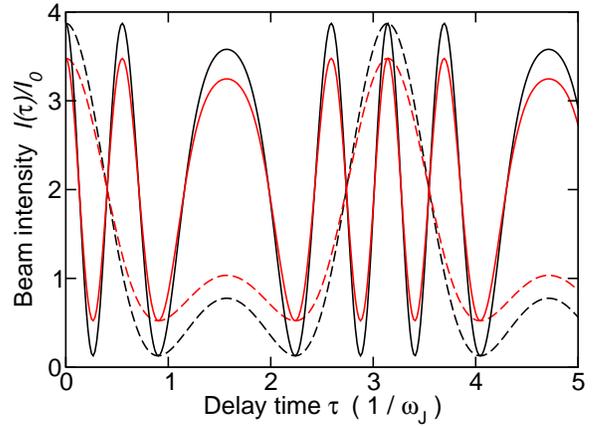}
\caption{(Color online).  Beam intensity $I(\tau)/I_0$
vs delay time $\tau$, for $\Delta U/\hbar\omega_J=$ 4, 12 (dashed
and solid lines, respectively) and $k_B T/\hbar\omega_J = $ 0, 2
(black [dark gray] and red [light gray] lines, respectively), with 
$\alpha(T=0)=$ 0.94. 
\label{fig3}}
\end{figure}

The maximum peak-to-valley ratio of fringes attainable for plasma oscillations
is limited by the condition that $\Delta U$ shifts the equilibrium
position $(k,\phi)_{\text{eq}}$ of the oscillator, 
as given in Eq.~(\ref{eq:displace}), slightly with respect to the origin:
$\Delta U\, \delta_J/(\hbar\omega_J)^2\ll 1$.
For example, by taking the values of $\omega_J$ and $\delta_J$
estimated in Sec.~\ref{s:estimate} and imposing the condition 
$\Delta U\, \delta_J/(\hbar\omega_J)^2=5\cdot 10^{-2}$,
one has $\Delta U / \hbar\omega_J \approx 12 $.
Figure \ref{fig3} displays $I(\tau )$ vs. $\tau$ for 
$\Delta U / \hbar\omega_J = 4, 12$ (dashed and solid lines, respectively).
In both cases the range of amplitude oscillations of $I$ is very close
to the ideal interval $\left[0,4I_0\right]$. Therefore, the intensity
oscillations should be easily detected, even at finite temperatures
(cf.~the black [dark gray] and red [light gray] lines, corrsponding
to $k_B T/\hbar\omega_J = $ 0, 2, respectively). 
For large values of $\Delta U / \hbar\omega_J$ higher overtones appear
in the oscillations of 
$I(\tau)$ (solid lines in Fig.~\ref{fig3}), in addition to the
fundamental frequency $\omega_J$, which is present for any finite value
of $\Delta U$ (cf.~dashed lines in Fig.~\ref{fig3}).

\section{Estimate of the visibility and plasma frequency}\label{s:estimate}

We assess the feasibility of the experiment
by estimating the parameters of Eq.~(\ref{interfe3}).
Both $\alpha$ and $\omega_J$ depend on $E_c$ and $\delta_J$.
We evaluate the latter by first solving the GP equation for
a two-dimensional harmonic trap within the Thomas-Fermi 
approximation,\cite{Pitaevskii}
and then by matching the wave functions of the two traps by
using the semiclassical method of Ref.~\onlinecite{Zapata}. 
The coupling constant $g=4\pi d e^2/\varepsilon_r$ 
appearing in the non-linear term of the GP equation,
multiplied by $n_s$, is the energy shift   
of an exciton added to 
a parallel plate capacitor with surface charge density $en_s$	
($\varepsilon_r$ is the quantum well dielectric 
constant).\cite{Butov-long} We obtain 
\begin{equation}
E_c=2\,\hbar\omega_0\,N_1^{-1/2}\left(\frac{d}{a_{\text{B}}}\right)^{1/2},
\end{equation}
where $\hbar\omega_0$ is the energy quantum of the trap 
and $a_{\text{B}}=\hbar^2\varepsilon_r/me^2$, as well as
\begin{equation}
\delta_J\sim\frac{u^2}{2^{1/3}\pi}
\left(\frac{a_{\text{B}}}{d}\right)^{2/3}\hbar\omega_0\,N_1^{-2/3}
\text{e}^{-S_0}[\tanh{S_0/2}]^{-1}\;.
\end{equation}
Here $u=0.397$, and 
$S_0\sim 2^{1/2}\pi(V_0-\mu)/\hbar\omega_0$
for a inter-trap quartic barrier of height $V_0$,
with $V_0-\mu\ll V_0$ [Fig.~\ref{fig1}(b)]. At density $n=2.5\cdot 10^{10}$
cm$^{-2}$, evaluated at the trap center, 
excitons are still weakly interacting
($na_{\text{B}}^2\sim 0.1$ with $a_{\text{B}}\approx$
20 nm). By taking GaAs parameters, $d=$ 12 nm, $N_1= 10^3$,
one has $\hbar\omega_0=$ 11 $\mu$eV, $\mu_1=$ 440 $\mu$eV,
and a condensate radius of $1.6$ $\mu$m. 
The barrier height $V_0$, as well as $S_0$, 
should be as low as possible.
For $S_0=1$  we obtain high visibility 
[$\alpha=0.94$ at $T=0$, cf.~Fig.~\ref{fig2}(b)],
as well as a plasma frequency $\omega_J/2\pi$ of 0.41 GHz, 
whose period ($\approx 2$ ns) is an order of magnitude shorter
than the exciton lifetime. Note that 
$\hbar\omega_J=$
1.7 $\mu$eV 
$\ll \hbar\omega_0$, hence the plasma oscillation
is decoupled from single-trap modes.\cite{Pitaevskii}
The temperature associated to $\hbar\omega_J$, $T=20$ mK, is 
very low but within experimental reach. 

The AC Josephson effect cannot be observed within our scheme.
In fact, for large values of $\Delta U$,
the term proportional to $\cos{\phi}$ appearing in the
Hamiltonian (\ref{H}) may be neglected in first approximation,
and the ground state wave function is a plane wave, $(2\pi)^{-1/2}
\exp{[i\bar{n}\phi]}$, where $\bar{n}$ is the integer closest to
$-\Delta U /E_c$. Since the probability density, $(2\pi)^{-1}$, is constant,
the phase is distributed randomly and
the visibility is zero. Therefore, the correction to $\alpha$ 
coming from the inclusion 
in the calculation of the term neglected 
in (\ref{H}) will be small and fragile 
against fluctuations.

\section{Conclusion}\label{s:conclusion}

In conclusion, 
exciton plasma oscillations may be measured
by the time correlation of photon emission from 
two sides of the Josephson junction
through electrical control of fringe visibility.
Our findings pave the way to 
the observation of the exciton Josephson effect.

We acknowledge stimulating conversations with L. Butov, A. Hammack,
S. Yang, I. Carusotto, V. Savona, C. Tejedor.
This work is supported by CNR Short Term Mobility Program 2008 and 
by the NSFPIF program.


\begin{thebibliography}{30}
\bibitem{Barone} A. Barone and G. Paterno, {\em Physics and Applications 
of the Josephson Effect} (Wiley, New York, 1982).
\bibitem{Helium} O. Avenel and E. Varoquaux, Phys.~Rev.~Lett.~{\bf 55,}
2704 (1985); S. V. Pereverzev, A. Loshak, S. Backhaus, J. C. Davis,
and R. E. Packard, Nature (London) {\bf 388,}
449 (1997); K. Sukhatme, Y. Mukharsky, T. Chui,
and D. Pearson, {\em ibid.} {\bf 411,} 280 (2001).
\bibitem{Atomic} F. S. Cataliotti, S. Burger, C. Fort,
P. Maddaloni, F. Minardi, A. Trombettoni, A. Smerzi,
and M. Inguscio, Science
{\bf 293,} 843 (2001); Y. Shin, M. Saba, T. A. Pasquini,
W. Ketterle, D. E. Pritchard, and A. E. Leanhardt, Phys.~Rev.~Lett.~{\bf 92,}
050405 (2004); M. Albiez, R. Gati, J. F\"olling, S. Hunsmann,
M. Cristiani, and M. K. Oberthaler, {\em ibid.} {\bf 95,}
010402 (2005); S. Levy, E. Lahoud, I. Shomroni, and J. Steinhauer,
Nature (London) {\bf 449,} 579 (2007).
\bibitem{Carusotto} M. Wouters and I. Carusotto, Phys.~Rev.~Lett.~{\bf 99,} 
140402 (2007). 
\bibitem{Sarchi} D. Sarchi, I. Carusotto, M. Wouters, and V. Savona, 
Phys.~Rev.~B {\bf 77,} 125324 (2008).
\bibitem{Shelykh} I. A. Shelykh, D. D. Solnyshkov, G. Pavlovic,
and G. Malpuech, Phys.~Rev.~B {\bf 78,} 041302(R) (2008).
\bibitem{Littlewood}
J. Keeling, F. M. Marchetti, M. H. Szyma\'nska, and P. B. Littlewood,
Semicond.~Sci.~Technol.~{\bf 22,} R1 (2007).
\bibitem{Oestreich96}
T. \"Ostreich, T. Portengen, and L. J. Sham,
Solid State Commun.~{\bf 100,} 325 (1996);
J. Fern\'andez-Rossier, C. Tejedor, and R. Merlin,
{\it ibid.}~{\bf 108,} 473 (1998);
A. Olaya-Castro, F. J. Rodr\'{\i}guez, L. Quiroga, and C. Tejedor,
Phys.~Rev.~Lett.~{\bf 87,} 246403 (2001).
\bibitem{Yang06}
S. Yang, A. T. Hammack, M. M. Fogler, L. V. Butov,
and A. C. Gossard, Phys.~Rev.~Lett.~{\bf 97,} 187402 (2006). 
\bibitem{Pitaevskii}
L. Pitaevskii and S. Stringari, {\em Bose-Einstein Condensation}
(Oxford University Press, Oxford, 2003).
\bibitem{Butov-long} 
L. V. Butov, J.~Phys.: Condens.~Matter {\bf 16,} R1577 (2004).
\bibitem{Hammack} A. T. Hammack, N. A. Gippius, S. Yang, G. O. Andreev,
L. V. Butov, M. Hanson, and A. C. Gossard, 
J.~Appl.~Phys.~{\bf 99,}
066104 (2006); A. A. High, E. E. Novitskaya, L. V. Butov,
M. Hanson, and A. C. Gossard, Science {\bf 321,} 229 (2008);
G. Chen, R. Rapaport, L. N. Pfeiffer, K. West, P. M. Platzman,
S. Simon, Z. V\"or\"os, and D. Snoke, Phys.~Rev.~B {\bf 74,} 045309 (2006).
\bibitem{RontaniPRL} Andreev-like phenomena related to
coherent exciton flow have been studied in
M. Rontani and L. J. Sham, Phys.~Rev.~Lett.~{\bf 94,}
186404 (2005); Solid State Commun.~{\bf 134,} 89 (2005).
\bibitem{Keldysh} L. V. Keldysh, in
{\em Bose-Einstein Condensation,} edited by A. Griffin, D. W. Snoke, S.
Stringari (Cambridge University Press, Cambridge, 1996), pp.~246-280.
\bibitem{SBGS} 
Only the relative phase between the two condensates has measurable effects,
see e.g. discussions in P. W. Anderson, Rev.~Mod.~Phys.~{\bf 38,}
298 (1966); A. J. Legget and F. Sols, Found. Phys. {\bf 21,} 353 (1991).
\bibitem{Balatsky}
A. V. Balatsky, Y. N. Joglekar, and P. B. Littlewood,
Phys.~Rev.~Lett.~{\bf 93,} 266801 (2004).
\bibitem{phase} Here we use the grand canonical formalism.
\bibitem{Smerzi} A. Smerzi, S. Fantoni, S. Giovanazzi,
and S. R. Shenoy, Phys.~Rev.~Lett.~{\bf 79,}
4950 (1997); S. Raghavan, A. Smerzi, S. Fantoni, and
S. R. Shenoy, Phys.~Rev.~A {\bf 59,} 620 (1999).
\bibitem{Zapata}
I. Zapata, F. Sols, and A. J. Leggett, Phys.~Rev.~A {\bf 57,} R28 (1998).
\bibitem{Oestreich} T. \"Ostreich and L. J. Sham, Phys.~Rev.~Lett.~{\bf 83,}
3510 (1999).
\bibitem{Glauber}
R. J. Glauber, in {\em Quantum Optics and Electronics,} C. DeWitt,
A. Blandin, C. Cohen-Tannoudji eds.~(Gordon and Breach, New York, 1965),
p.~63.
\bibitem{notespectral}
Another possibility to measure the exciton Josephson effect would be 
the observation of Bogoliubov excitations from the spectral properties 
of the emitted light.\cite{Sarchi} A drawback of this idea is that
spectral properties are not unambiguosly linked to the condensed phase.
\bibitem{Zimmermann}
R. Zimmermann, Phys.~Stat.~Sol.~(b) {\bf 243,} 2358 (2006).
\bibitem{Andrews} M. R. Andrews, C. G. Townsend, H.-J. Miesner,
D. S. Durfee, D. M. Kurn, and W. Ketterle, Science {\bf 275,} 637 (1997).
\bibitem{Pita} 
L. Pitaevskii and S. Stringari, Phys.~Rev.~Lett.~{\bf 87,} 180402 (2001).
\end{thebibliography}
\end{document}